\documentclass[10pt]{iopart}

\usepackage{iopams}  
\usepackage{graphicx}
\graphicspath{{figures/}}
\usepackage{bm}
\usepackage{hyperref}
\usepackage{lineno}

\begin{document}

\title[]{Vlasov-Poisson simulation study of phase-space hole coalescence in a cylindrically wave-guided plasma}

\author{Allen Lobo \& Vinod Kumar Sayal}

\address{Dept. of Physics, Sikkim Manipal Institute of Technology, Sikkim Manipal University, Sikkim 737 136, India.}
\ead{allen.e.lobo@outlook.com}

\begin{abstract}
In this work, coalescence of phase-space holes of collision-less, one-dimensional plasmas is studied using kinetic simulation techniques. Phase-space holes are well-known Bernstein-Greene-Kruskal waves known for exhibiting coalescence, are numerically simulated and their coalescence is observed. Relations between the hole speed, potential, phase-space vorticity and phase-space depth are then obtained using the simulation data. This study involves the study of electron phase-space hole coalescence in a cylindrically wave-guided plasma. Using the recently developed phase-space hydrodynamic analogy, it is shown that the coalescence phenomena can be explored in terms of the fluid-analogous vortical nature of the phase-space holes. Coalescence occurs due to the interaction of the phase-space velocity fields associated with these phase-space vortices. Results obtained from the study describes various parametric relations between the coalesced hole characteristics and the characteristics of the colliding holes.
\end{abstract}
\vspace{2pc}
\noindent{\it Keywords}: Phase-space hydrodynamic theory, Phase-space holes, Hole coalescence

\section{Introduction}
As a medium, plasma is home to a spectrum of nonlinear wave phenomena, which arise due to numerous instabilities and excitation techniques. Although these waves have been addressed ubiquitously in the existing literature, they still remain sufficiently unexplored, inciting one to analyse them using various analytical, simulation-based or experimental techniques. Among these nonlinear waves, phase-space holes represent one of the most interesting modes which the kinetic theory of collisionless plasma addresses. It is known that beam instabilities in plasmas generate phase-space holes due to particle trapping in their nonlinear evolution stages. They have been observed in countless numerical simulations \cite{Roberts1967NonlinearInstability, morse1969one, Saeki1979, Lynov1979, Lynov1980, Guio2003, Lynov1985Phase-SpaceHoles, turikov1978computer, Eliasson2004ProductionHole, Eliasson2004DynamicsPlasma}, laboratory experiments \cite{Saeki1979, Lynov1979} and space plasma observations \cite{Wang2022MultisatelliteSheet, Lotekar2020MultisatelliteInstability, Aravindakshan2021StructuralPlasma, Vasko2020OnShock,Wang2017, Pickett2008FurtheringTheory}, in which they appear as Debye-scaled, bipolar electrostatic solitary structures travelling with speeds comparable to the thermal speed of the plasma species. Holes present themselves as unique and well-identifiable phase-space particle density depletion zones, surrounded by regions of high particle densities. Their characteristic inverted-bell shaped, solitary waveforms of electrostatic potentials are positive in the case of electron holes and negative in the case of ion holes. Phase-space holes are Bernstein-Greene-Kruskal modes \cite{Bernstein1957}, and exhibit some characteristic kinematics and evolutionary behavior, including remarkable stability during its solitary wave propagation \cite{Saeki1979, schamel1986electron} and particle reflection and self-accelerations \cite{Eliasson2004ProductionHole, Eliasson2004DynamicsPlasma, Eliasson2006}.

Among these exhibited characteristic behaviors, phase-space holes also exhibit the phenomena of coalescence. This was first observed in two-stream plasmas \cite{Roberts1967NonlinearInstability, morse1969one, Berk1970PhaseObservations} in which the instabilities lead to the formation of electron phase-space holes, which emerge and coalesce with each other until a single, stable hole remains. A more detailed simulation-based study of electron hole coalescence in a cylindrical wave-guided plasma \cite{Lynov1980} revealed that electron holes coalesce when their relative velocities are comparatively small. In the case of large velocity difference, the phase-space holes simply pass through each other, without affecting each other's propagation characteristics. Later works of Eliasson and Shukla \cite{Eliasson2006} showed via simulation studies of pair-plasmas and relativistic plasmas that this coalescence is an inelastic collision of the two holes which results in the formation of a stable hole which lasts throughout the simulation. 

Both electron and ion phase-space holes resemble in nature to fluid vortices due to the trajectories of the trapped particles. In their respective phase-spaces, the trapped electrons (or ions) in the electron hole (or ion hole) region oscillate in rotational trajectories, with bounce frequencies which are constant, irrespective to their distance from the gyration centre, hence exhibiting this similarity with conventional two-dimensional fluid vortices. This comparison is also supported by the fact that the governing Vlasov-Poisson equations are in-fact analogous to the continuity equation of an in-compressible fluid \cite{Berk1970PhaseObservations, schamel1986electron, schamel2012cnoidal, Luque2005ElectrostaticSystems}. This analogy is further extended by the presence of vorticity fields and fluid-analogous vortex identification techniques in the phase-space which conveniently recognize phase-space holes as vortices, therefore exhibiting and further supporting their vortical nature, as shown by Lobo and Sayal \cite{Lobo2024TheoryPlasmas}. In this recent paper, the fluid-like phase-space hydrodynamic analysis of the phase-space hole structure makes it convenient to analytically study the structure of phase-space hole by employing suitable conditions of the system, such as its thermodynamic homogeneity and isotropic nature.

Inspired by the phase-space hydrodynamic approach, this article aims to investigate the phenomena of phase-space hole coalescence in terms of the existing fluid-analogous properties of the phase-space. We deal with the case of an electron-ion plasmas. We initially investigate the electron hole coalescence phenomena in the two-stream electrostatic continuous plasma \cite{morse1969one, Roberts1967NonlinearInstability}. We find that this coalescence interaction is induced by the interactions of the velocity fields of the individual phase-space vortices. We then study the coalescence of electron holes in a cylindrical wave-guided set-up, similar to the simulated experiment by Lynov et al. \cite{Lynov1980}, in which these phase-space structures exhibit stability during their propagation. We similarly simulate and analyze ion hole coalescence in the wave-guided plasma using similar techniques as conducted by Lobo and Sayal \cite{Lobo2024AccelerationPlasma}. In either case, we investigate the interactions of the individual potential and other dynamical fields of each hole and discuss the observed relations which result from these interactions. This study helps to understand the phase-space hydrodynamics associated with the coalescence of phase-space vortices.

\section{Methods}
\subsection{Theory and governing equations}
As defined initially by Lobo and Sayal \cite{Lobo2024TheoryPlasmas}, the phase-space vorticity field $\xi$ describes the vortical contortions in the phase-space, as in the case of a conventional fluid wherein the fluid vorticity field describes the relative deformations in local fluid regions. Analogous to conventional fluid dynamics, phase-space vorticity field is defined as the curl of the fluid velocity field $(\bm{\mathcal{V}})$. 
\begin{equation}
    \bm{\mathcal{V}}(x,\tau v_x) = v_x \hat{x} - \frac{\tau q}{m}\frac{\partial \phi(x)}{\partial x}\hat{v}_x, \quad \tau = \omega_{p}^{-1} = \sqrt{\frac{\varepsilon_0 m}{n_0 q^2}}.
\end{equation}
Here, $\tau= \omega_p^{-1}$ represents the inverse of plasma frequency, and has been attached to the velocity space $(\hat{v}_x)$ in order to equate the position and velocity dimensions. As is the case of the usual notations, $x$ represents position space, $v_x$ the velocity space and $q/m$ represents the charge-to-mass ratio of the species. Similarly, $n_0$ and $\varepsilon_0$ respectively represent the uniform plasma number density and free-space electrostatic permittivity. The scalar potential $\phi(x)$ can be used to determine the electric field $E(x)$ and hence, the phase-space fluid velocity along the velocity space. Upon using the phase-space derivative operator $\bm{\nabla}$ as,
\begin{equation}
    \bm{\nabla} = \frac{\partial}{\partial x}\hat{x} + \frac{1}{\tau}\frac{\partial}{\partial v_x}\hat{v}_x,
\end{equation}
the phase-space vorticity field can be deduced and is found to be a function of the spatial charge density of the system, 
\begin{equation}\label{vorticity_field}
    \xi = \bm{\nabla} \times \bm{\mathcal{V}} = \omega_{p}\left(\frac{\rho(x)}{qn_0} -1\right).
\end{equation}
The result presented in equation (\ref{vorticity_field}) is particularly interesting \cite{Lobo2024TheoryPlasmas}. Firstly, the dependence of the phase-space vorticity on the spatial charge density implies that contortions in phase-space plane occur in the case of collisionless, electrostatic plasmas, due to localized charge accumulations. Specifically, according to Lugt \cite{Lugt1979} and Tian et al.\cite{Tian2018a}, a region of locally concentrated vorticity field in a fluid directly corresponds to high contortions, which usually become vortices. The phase-space identification criteria which recognizes phase-space holes \cite{Lobo2024TheoryPlasmas}, identifies holes as vortices based on concentration of vorticity fields along position space. However, the flow of the phase-space fluid is already in the clockwise direction around the $v_x=0$ axis, as is its innate property. As per Tian et al. \cite{Tian2018a}, vorticity causes local deformation of the flow field into a rotational flow. For the formation of a vortex in phase-space, this deformation is needed only across the position space. Hence, vorticity field concentrations along the position space are sufficient criteria for phase-space vortices. 

Hence, in this phase-space hydrodynamic analogy, a phase-space hole is a vortex in the phase-space formed due locally concentrated charge densities which contribute towards the formation of locally concentrated vorticity field, as is discussed by Lobo and Sayal \cite{Lobo2024TheoryPlasmas}. Secondly, it can be seen that the phase-space vorticity field magnitude $|\xi|$ remains even in regions with uniform charge distributions and no net charge density. This implies that the phase-space itself has an innate vorticity field due to which it exhibits sheering flow, specifically along the position axis, which is well known. The phase-space flow also exhibits a no-flux condition \cite{Lobo2024TheoryPlasmas},
\begin{equation}
    \bm{\nabla}\cdot\bm{\mathcal{V}}=0.
\end{equation}
This means that the presence of a spatial charge density does not produce the vorticity field, but enhances it, causing a rotational deformation of the hydrodynamic phase-space flow. The flow of the phase-space is by nature solenoidal in the collisionless case \cite{Lobo2024TheoryPlasmas, Lobo2024ASystems}. Therefore, a contribution to the phase-space vorticity field which results in the formation and dynamics of phase-space vortices, directly results from the dynamics of the spatial charge densities. In other words, it is not the vorticity field itself, but a change in the vorticity field $|\delta \xi| = |\delta \rho|$ which contributes to vortical formation and dynamics in the plasma phase-space. 

However, it is important to note a key role of the phase-space vorticity field. Even though the spatial charge density and the phase-space vorticity field analysis bears a linear relationship, it is the presence of the phase-space vorticity field which inadvertently leads to the presence of phase-space vortices. Therefore, even though quantitatively the phase-space vorticity field is dependent directly on the spatial charge density, and a measurement of the latter directly implies the measurement of the former, it is qualitatively the presence of the phase-space vorticity field which describes presence of a phase-space hole \cite{Lugt1979, Tian2018a}. In the succeeding analyses, we therefore resort to the study of spatial charge density, and correlate to it the vorticity field analysis using equation (\ref{vorticity_field}).

The collisionless plasma kinetics is collectively governed by the Vlasov-Poisson equations system,
\begin{equation}\label{vlasov}
    \frac{\partial f_i}{\partial t}+v_x\frac{\partial f_i}{\partial x}-\frac{q}{m}\frac{\partial \phi}{\partial x}\frac{\partial f_i}{\partial v_x}=0,
\end{equation}
\begin{equation}\label{Poisson}
    \frac{\partial^2\phi}{\partial x^2} = -\frac{n_0}{\varepsilon_0}\sum_{i=1}^n \int_{-\infty}^{\infty}q_i f_i(x,v_x)dv_x.
\end{equation}
Here, $f_i(x,v_x)$ represents the phase-space density of the $i$-th species of the system. Specific to the case studied, the phase-space density and the associated plasma parameters can be modified accordingly. For the case of electron hole generation and propagation, initial distribution of the electrons $(f_e)$ are assumed to be Maxwellian in nature:
\begin{equation}\label{f_maxwellian}
    f_e(x,v_x) = n_0\sqrt{\frac{1}{\pi}}\exp\left(- \frac{m_e v_x^2}{2K_BT_e}\right).
\end{equation}
Here, $m_e$ represents the electron mass, $K_B$ and $T_e$ respectively the Boltzmann's constant and electron temperature. Ions are assumed to remain stationary and uniform along space, with a constant spatial number density $n_0$. Ions therefore form a singly ionised, neutralising background. Periodic boundary conditions are employed such that, for the length $L$ of the plasma column,
\begin{equation}\label{periodic_boundaries}
    f_e(x>L,v_x) \Rightarrow f_e( x-L,v_x),\quad f_e(x<0,v_x) \Rightarrow f_e( L+x,v_x) .
\end{equation}
 For the case of a cylindrically wave-guided plasma, only the lowest radial eigenmode of the transverse Laplacian eigenstate is used. Hence, the Poisson equation is modified as,
 \begin{equation}\label{radial_poisson}
     \frac{\partial ^2\phi}{\partial x^2} - k_\perp ^2\phi(x) =  -\frac{n_0}{\varepsilon_0}\left(1-\int_{-\infty}^{\infty}q_i f_i(x,v_x)dv_x\right).
 \end{equation}
 For the case of ion hole studies, we consider an electron-ion plasma in which ion holes follow an initial Maxwellian distribution similar to equation (\ref{f_maxwellian}). However, the electrons are assumed to be thermalised, such that their spatial density becomes:
 \begin{equation}\label{thermal_electrons}
     n_e = n_0 \exp\left(\frac{e\phi(x)}{K_BT_e}\right).
 \end{equation}
 Electron to ion temperature ratio is taken to be $T_e/T_i = 100.0$. In the case of an infinite electron two-stream plasma, the electron distribution obtains the form of a modified Maxwellian case, with the phase-space density of the form:
 \begin{equation}\label{two_stream_distribution}
      f_e(x,v_x) = \frac{m_e n_0}{2K_BT_e}\sqrt{\frac{1}{\pi}}\exp\left(- \frac{m_e v_x^2}{2K_BT_e}\right)\cdot v_x^2.
 \end{equation}
 
 The Vlasov-Poisson system and other equations presented in equations (\ref{vlasov}-\ref{thermal_electrons}) are normalised appropriately by expressing the position $x$ in terms of the electron (ion in case of ion hole simulations) Debye length $\lambda_{De} = \sqrt{\varepsilon_0K_BT_e/n_0 e^2}$, velocity $v_x$ in terms of the thermal velocity $v_{Te}=\sqrt{2K_BT_e/m_e}$, time in units of inverse plasma frequency $\omega_{pe}^{-1} = \sqrt{\varepsilon_0 m_e/n_0e^2}$ and potential in units of the thermal potential $K_BT_ee^{-1}$. 
 
 \subsection{Simulation techniques}
 In order to numerically integrate the Vlasov equation to simulate the time-evolution of the plasma phase-space, we employ the well-known finite-difference scheme \cite{Cheng1976}, which has been shown to be a high-order, highly accurate, semi-Lagrangian numerical integration scheme for the Vlasov equation \cite{Filbet2001}. The finite-difference scheme employs an interpolating technique to numerically evolve the phase-space density by shifting it to its advective neighborhood points in phase-space, for half time-steps, following a leap-frog pattern. In this work, we have used the cubic spline interpolation technique, which is known to be a fourth order interpolation scheme. The Poisson equation is solved by using the inverse matrix method \cite{Ndayisenga2022Finite-differenceProblem}. For the numerical simulation, the phase-space is represented by a space-velocity grid of $2^6\times 2^8$ points along position and velocity, respectively. The Courant-Friedrichs-Lewy condition \cite{Courant1928} is employed to reduce the numerical dissipation, with the Courant number as $1$. Normalised forms of the Vlasov-Poisson systems are numerically integrated at each time-step. The results and associated discussions are presented in the succeeding sections.

\section{Simulation results and discussions}
In order to simulate phase-space hole coalescence, we employ suitable perturbation techniques in the form of particle density or potential perturbations. We study the coalescence phenomena in terms of the phase-space velocity field $(\bm{\mathcal{V}})$, vorticity field $\xi$, potential $(\phi)$ and hole speeds $(M)$, and hole depth in phase-space $(\Gamma)$. We consider initially the study of electron phase-space hole coalescence in the two-stream case, as was initially shown by Morse and Nielson \cite{morse1969one} . This is followed by the cylindrically wave-guided plasma study. 

\subsection{Two-stream electron hole coalescence}

It is known that instabilities in two-stream plasmas generate phase-space holes in their nonlinear evolution stages. It has been shown in various numerical simulations and recent space plasma observations \cite{morse1969one, Berk1970PhaseObservations, hutch2017, Dubinov2020NonlinearCavity, Wu2012TheInstability} that instabilities produced in two-stream collisionless plasmas form phase-space holes which remain stable in one-dimensional observations, but become unstable in two or three dimensional cases due to increasing transverse instabilities. These phase-space holes exhibit coalescence during their evolutionary stages, until a single, stationary, stable hole exists. This was shown in the numerical simulations of Morse et al. \cite{morse1969one}.

In this study, we simulate the excitation of electron phase-space holes in two-stream plasmas, following an initial phase-space distribution represented by equation (\ref{two_stream_distribution}). We use an initial spatial particle density perturbation of the hyperbolic-secant form, in a spatially symmetric system, such that $x\in [-L,L]$. Here we take the length equal to $2L$, which is from $-L$ to $L$, where $L=100.0\lambda_{De}$. The boundary conditions presented in equation (\ref{periodic_boundaries}) are therefore adjusted accordingly. Electron hole formation begins at the nonlinear trapping stage of the system. Hole coalescence initiates in the later stages of the plasma evolution. 

\begin{figure}
    \centering
    \includegraphics[width=1\linewidth]{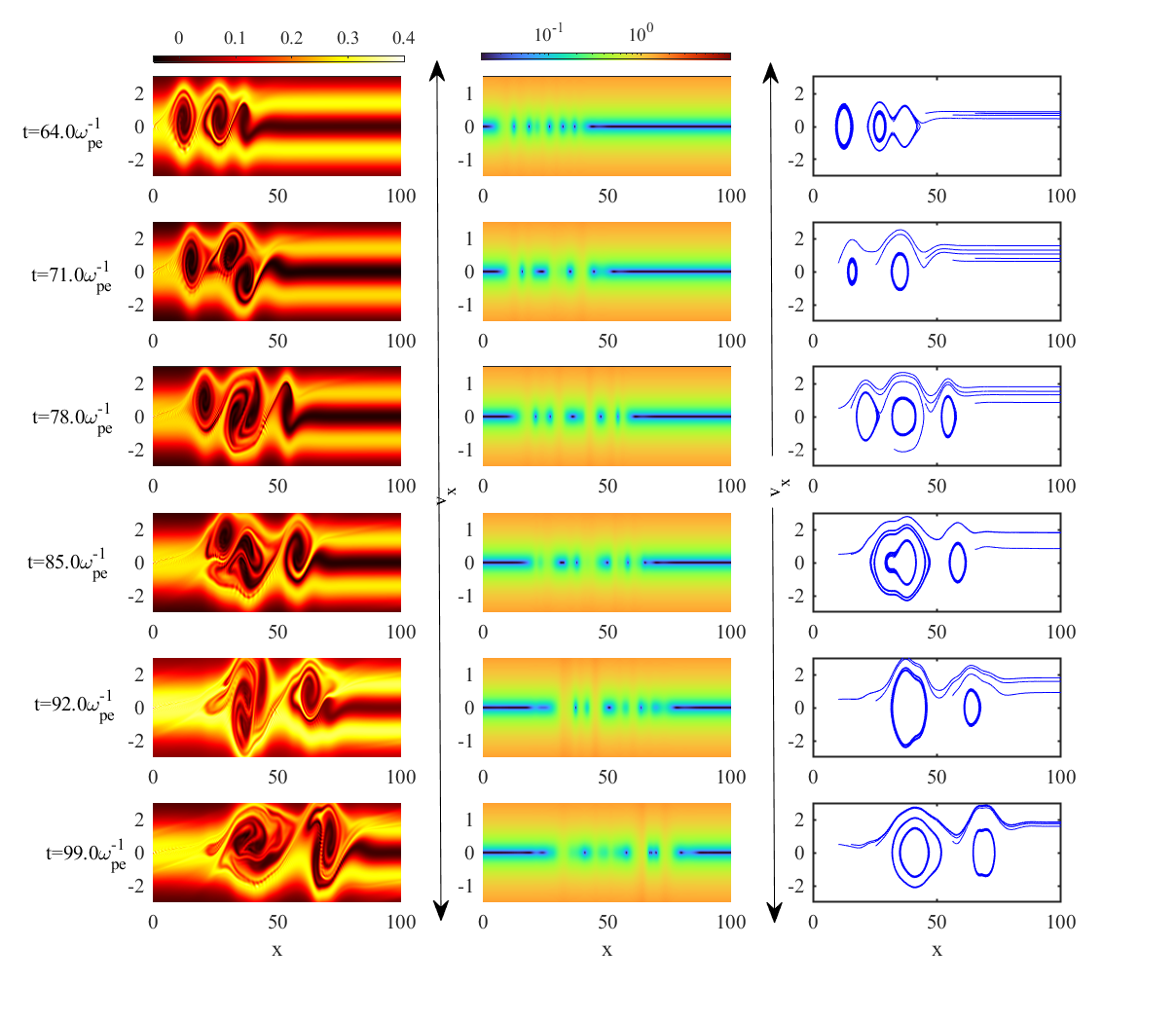}
    \caption{Electron hole coalescence in the two-stream plasma. (Left column) Phase-space density portrait (in linear scale), (middle column) phase-space velocity field magnitude $|\bm{\mathcal{V}}|$ portrait (in logarithmic scale) and (right column) streamlines of phase-space volume elements showing the interactions of velocity fields of the holes during the hole coalescence. It can be seen that holes (phase-space vortices) coalesce due to interactions and merging of their phase-space velocity fields. Each streamline shows particle motion in clockwise direction.}
    \label{fig:v_mag_2str}
\end{figure}

We found from our simulation study of the two-stream holes that the phenomena of hole coalescence can be attributed to the interactions of individual phase-space velocity fields $\bm{\mathcal{V}}$ of each phase-space hole (see figure \ref{fig:v_mag_2str}). Specifically, the electron holes coalesce by the merging (fusion) of the central, non-rotating cores of the phase-space vortices. In the phase-space hydrodynamic analogy, each electron hole is a phase-space vortex with the phase-space fluid orbiting around a central, stationary core. The core of these vortices refer to the point in the phase-space where $|\bm{\mathcal{V}}|=0$, located at the central minima of the hole phase-space density. This core does not rotate, but moves along the length of the plasma column with the hole speed $M$. When two phase-space holes are at sufficiently close distance such that their orbiting phase-space fluid velocity fields interact, their vortical cores begin to fuse together. This causes their individual velocity fields to merge and collectively behave as a single vortical structure, which marks their fusion. Figure \ref{fig:v_mag_2str} (right column) shows this modulation of the trapped electron orbits which is marked by the velocity fields of the electron phase-space vortices, the magnitudes of which are shown in figure \ref{fig:v_mag_2str} (middle column).

\subsection{Hole coalescence in cylindrically wave-guided plasma}

\begin{figure}[!ht]
    \centering
    \includegraphics[width=1\linewidth]{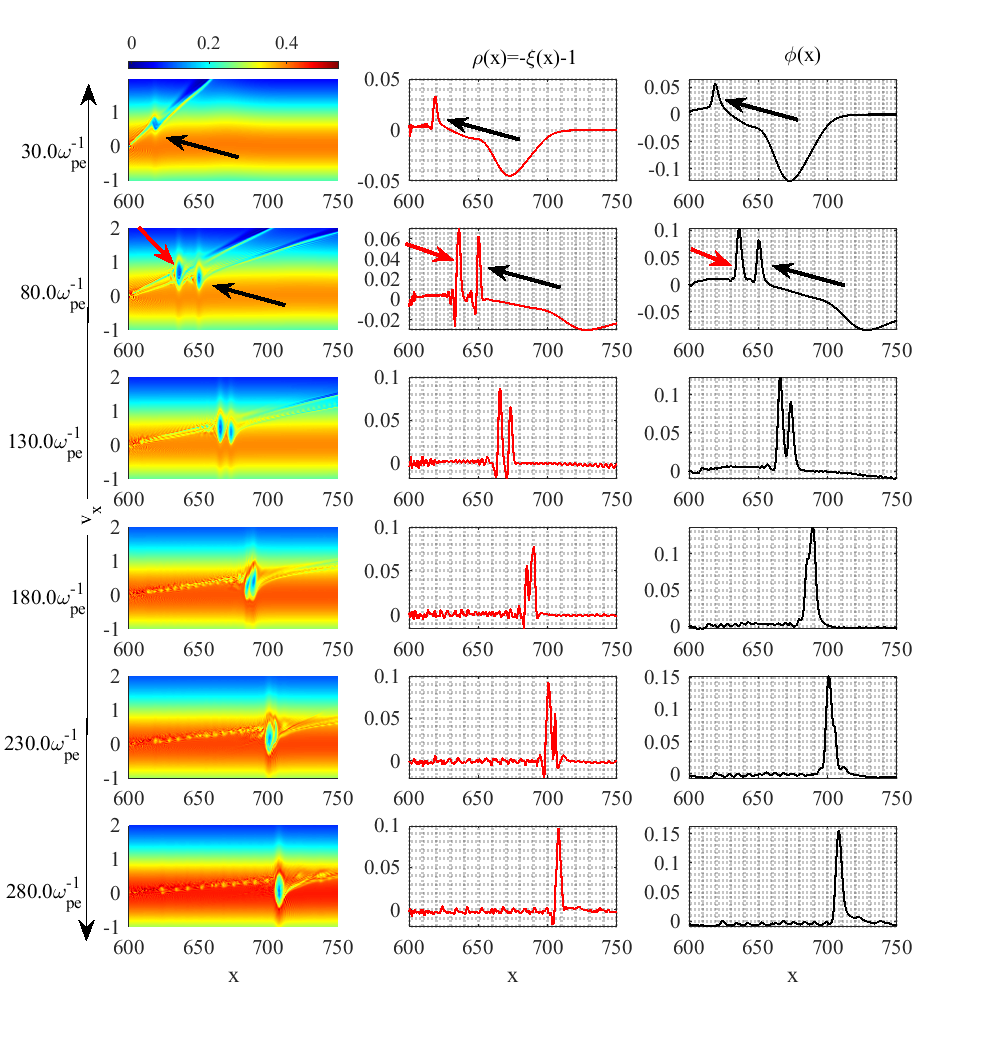}
    \caption{Excitation and coalescence of solitary electron holes in a cylindrically wave-guided plasma. Observations of phase-space density (left column), spatial charge density (middle column) and potential (right column) at different time-steps during the coalescence. Excitation produced by pulse amplitude $A=2.5K_BT_ee^{-1}$ for the slow hole (pointed by black arrow) and $A=3K_BT_ee^{-1}$ for the fast hole (pointed using red arrow), such that $\mathcal{K}=1.2$.}
    \label{fig:elec_hole_q_m}
\end{figure}

\begin{figure}[!ht]
    \centering
    \includegraphics[width=1\linewidth]{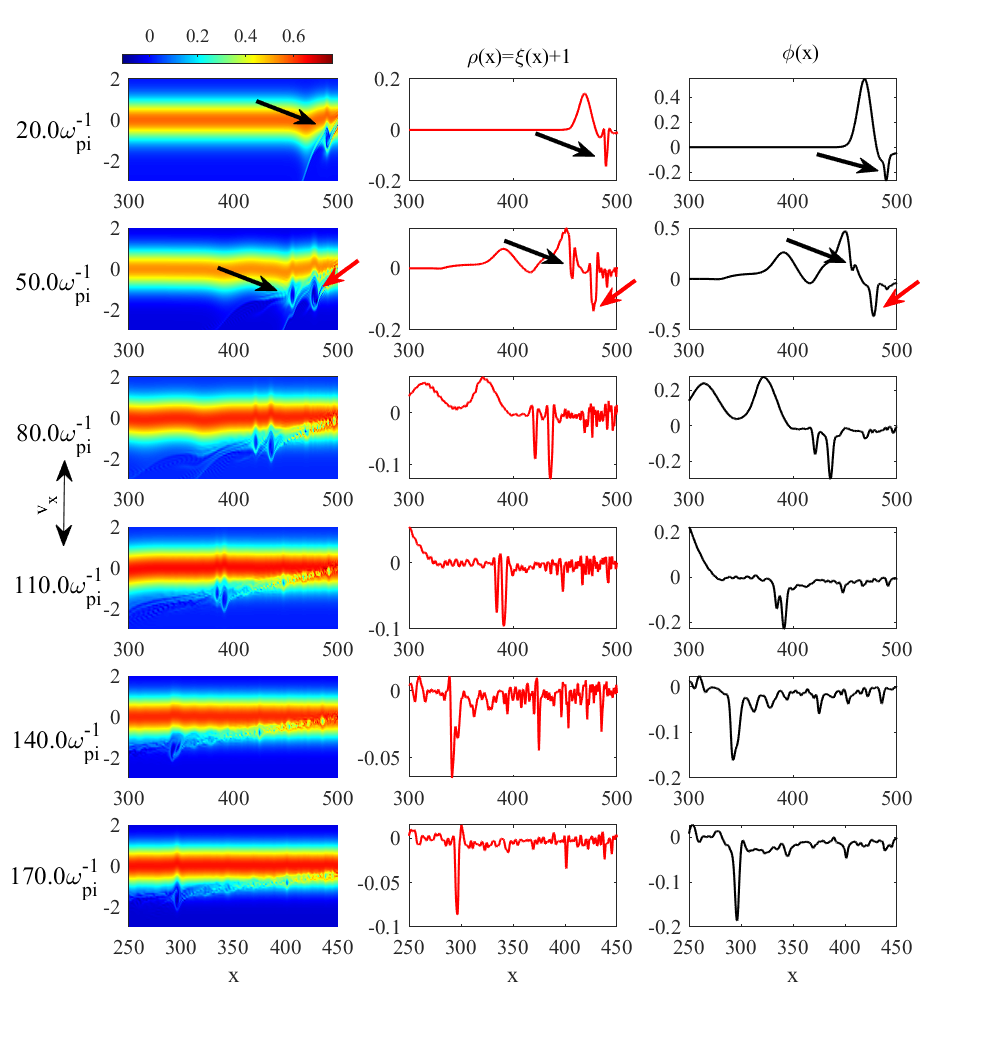}
    \caption{Excitation and coalescence of solitary ion holes in a cylindrically wave-guided plasma. Observations of phase-space density (left column), spatial charge density (middle column) and potential (right column) at different time-steps during the coalescence. For the excitation of plasma, we take $A = 3.0K_BT_ie^{-1}$ for the slow hole (indicated by the black arrow) and $\mathcal{K} = 1.5$ for the fast hole (indicated by the red arrow).}
    \label{fig:ion_hole_q_m}
\end{figure}

In this section, we study the hole coalescence in a radially confined, cylindrically wave-guided plasma by employing a kinetic Vlasov simulation of the experiment performed by Turikov \cite{turikov1978computer}, Saeki et al. \cite{Saeki1979}, Lynov et al. \cite{Lynov1980} and Lobo and Sayal \cite{Lobo2024AccelerationPlasma}. We present the simulation observations for both electron and ion hole cases. Since the results of our study are similar for both cases, with the ion hole potential and charge densities having negative polarities (opposite to the electron hole case, which is well-expected), we omit discussions on the ion hole case. The electron phase-space hole excitation and coalescence is shown in fig. \ref{fig:elec_hole_q_m} and the ion hole excitation and coalescence is shown in fig. \ref{fig:ion_hole_q_m}. In either case, the plasma column length $L=1200\lambda_{De (or) Di}$. The radius of the cylindrical wave-guide is taken to be $4\lambda_{De}$ in case of electron hole excitation and $5\lambda_{Di}$ in the case of ion holes. 

In order to excite the plasma, a step potential is used, similar to the excitation technique described by Turikov \cite{turikov1978computer}. The amplitude of the pulse is varied, keeping its width a constant. This excitation technique of the plasma column is well-known to form a solitary hole. A second pulse of increased amplitude is introduced again in order to produce a second electron hole, moving with a larger speed. The second excitation potential is taken to be $\mathcal{K}$ times the first excitation amplitude, where $\mathcal{K} \in [1,1.5]$. This method of excitation causes overtaking collisions of the two holes. Depending on the relative speeds of the two electron holes, hole coalescence is then observed. Specifically, we notice that when $\mathcal{K}$ is increased beyond $1.5$, hole collisions do not result in coalescence. The simulation snapshots are shown in the figures \ref{fig:elec_hole_q_m} and \ref{fig:ion_hole_q_m}.

\begin{figure}
    \centering
    \includegraphics[width=0.5\linewidth]{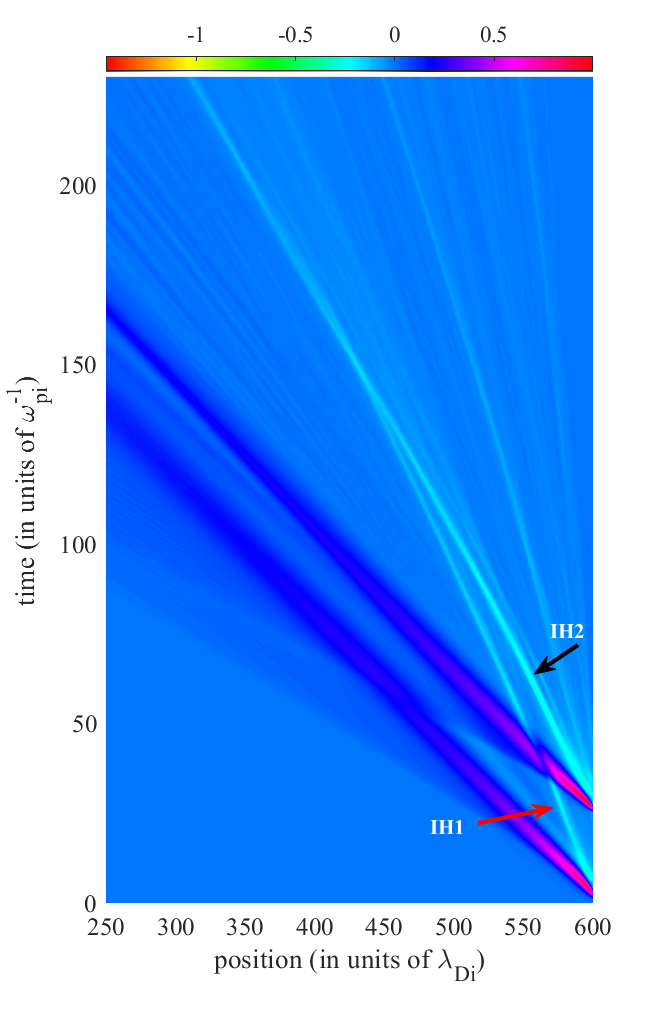}
    \caption{Spatio-temporal portrait of ion hole coalescence in a cylindrically wave-guided set-up. The excitation, propagation and coalescence of the two holes (marked as IH1 and IH2) can be seen. The acceleration of ion holes during their propagation, as initially shown by Lobo and Sayal \cite{Lobo2024AccelerationPlasma} can also be seen. Excitation of ion holes initiated by using pulse amplitude $A=4$ and $\mathcal{K}=1.5$. }
    \label{fig:ih_coal_cont}
\end{figure}

Observations of spatial charge densities $\rho(x)$, hole potential $\phi(x)$ and hole phase-space densities are reported for the coalescing holes, before, during and after their coalescence into a single hole. The stability of the new hole is observed for sufficiently large time, till the traveling waves reach the axial boundaries of the simulation, after which the observations are considered erroneous. We also observe the acceleration of the ion holes in the wave-guided plasma during their propagation, as initially reported by Lobo and Sayal \cite{Lobo2024AccelerationPlasma}. This can be noticed as the bending of the hole paths as they propagate towards the time axis in figure \ref{fig:ih_coal_cont}, which also shows the phase-shift occurring during the collision of the ion hole and the ion KdV soliton (pink, negative pulse).

In order to observe the physical characteristics of the coalesced electron holes and their dependence on the propagation characteristics of the solitary holes before their coalescence, we vary the perturbation amplitude and observe the corresponding changes in our readings. Specifically, we increase the perturbation amplitude, due to which the potential amplitude $(\phi_0)$, speed $(M)$ and charge density amplitude $(\rho_0)$ of each colliding solitary hole increase. We then observe coalescence between the holes and note the characteristics of the coalesced hole. Since these characteristics depend solely on the characteristics of the individual holes before their collision, we measure the coalesced hole characteristics against the changing properties of the individual holes.

\begin{figure}[!ht]
    \centering
    \includegraphics[width=0.45\linewidth]{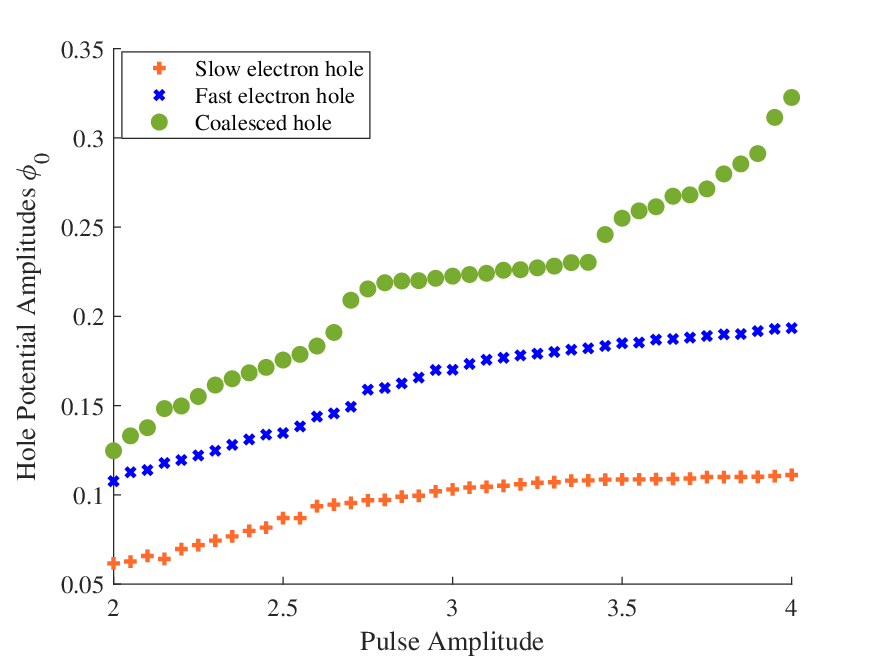}
    \includegraphics[width=0.45\linewidth]{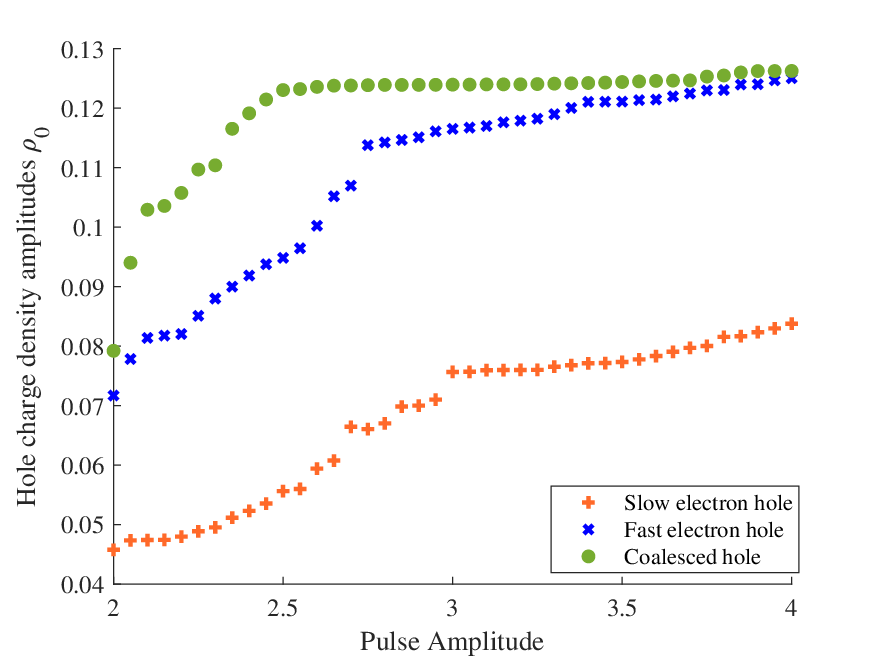}
    \includegraphics[width=0.45\linewidth]{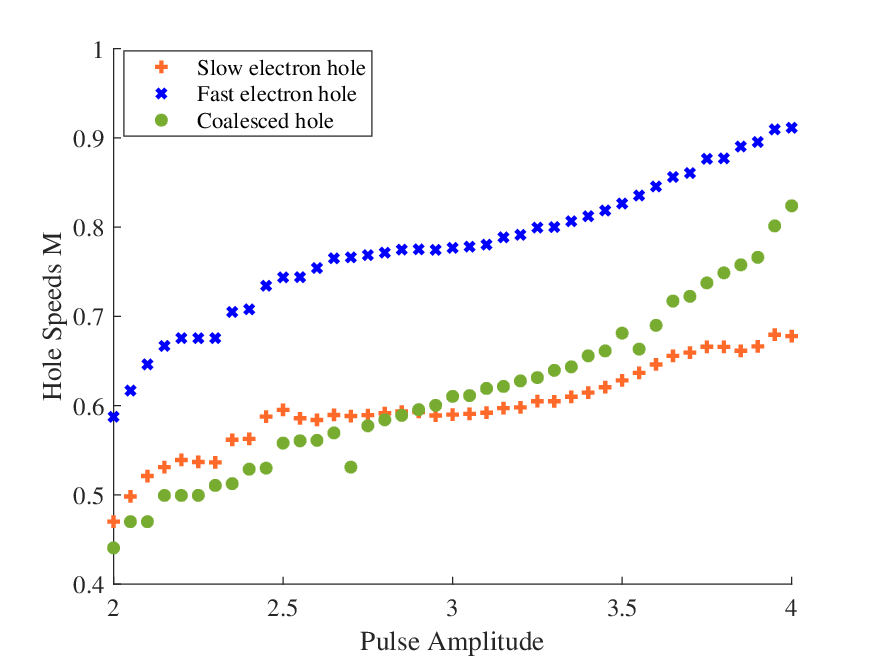}
     \includegraphics[width=0.45\linewidth]{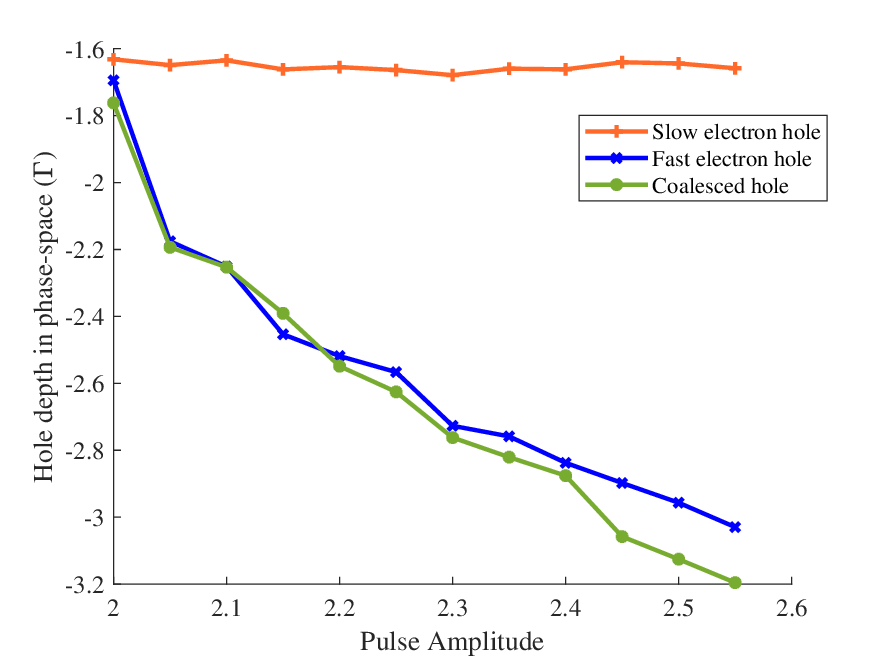}
    \caption{Variation of hole (top left) potential amplitudes, (top right) charge density amplitudes, (bottom left) hole speeds and (bottom right) hole phase-space depths with increasing excitation potential amplitude in the cylindrically wave-guided plasma set-up. During the studies (top-left, top-right and bottom-left), we take $\mathcal{K}=1.5$ for the excitation of the faster hole. For the hole depth measurement (bottom right), we vary the amplitude of the second potential using $\mathcal{K}\in [1.01, 1.3]$. }
    \label{fig:electron_hole_qm_potential_vs_amp}
\end{figure}

It can be seen from figure \ref{fig:electron_hole_qm_potential_vs_amp} (top left) that as the perturbation amplitude increases, the solitary electron hole potential amplitudes also increase, as is expected. It can be seen that the coalesced electron hole potential amplitude increases as the amplitude of the individual hole increases. Since this hole potential depends on the individual coalescing holes and not the perturbation amplitude, this observation depicts a direct dependence of the coalesced hole potential on the amplitudes of the individual coalescing (slow and fast) holes. However, this dependence is clearly not linear, since the coalesced hole amplitude is lesser than the linearly superimposed sum of the two individual hole amplitudes.

A similar relation is noticeable in the case of hole charge density amplitude, and accordingly the vorticity field amplitude. It can be seen in figure \ref{fig:electron_hole_qm_potential_vs_amp} (top right) that the individual charge density amplitudes of both slow and fast electron holes in the plasma increase upon increasing the perturbation amplitude. This is expected since their potential amplitudes also increase, and from equation (\ref{vorticity_field}) it is clear that vorticity $\xi(x)$ is a function of the spatial charge density $\rho(x)$ which inadvertently is dependent on the potential field $\phi(x)$ from the Poisson's equation (\ref{radial_poisson}). Figure \ref{fig:electron_hole_qm_potential_vs_amp} (bottom left), in the same way, showcases the hole speeds as a function of the excitation potentials. It is well-known that the hole coalescence is an inelastic collision between the two colliding holes \cite{Lynov1980, Eliasson2006}, due to which the resultant coalesced hole speed is reduced. In order to investigate the effect of hole coalescence on the phase-space depth $(\Gamma)$ of the phase-space vortices, we induce collisions between holes in a slightly different way. The phase-space hole depth is calculated as the difference between the natural logarithms of the phase-space densities at the hole core (minima) and boundary, where $v_x = \sqrt{\phi_0}$ \cite{Lobo2024TheoryPlasmas}, and is a negative value. The first (slow) electron hole is generated using a constant perturbation amplitude $(A)$. The second, faster electron hole is then excited by using the perturbation amplitude $\mathcal{K}A$, where $\mathcal{K}$ is varied from $1.01$ to $1.3$. In each case, the phase-space depths of each hole is measured, along-with the depth of the coalesced hole. Figure \ref{fig:electron_hole_qm_potential_vs_amp} (bottom right) presents our findings. It can be seen that the coalesced hole depth increases with the increase in hole depth of the colliding hole.

\begin{figure}\centering
    \includegraphics[width=0.48\linewidth]{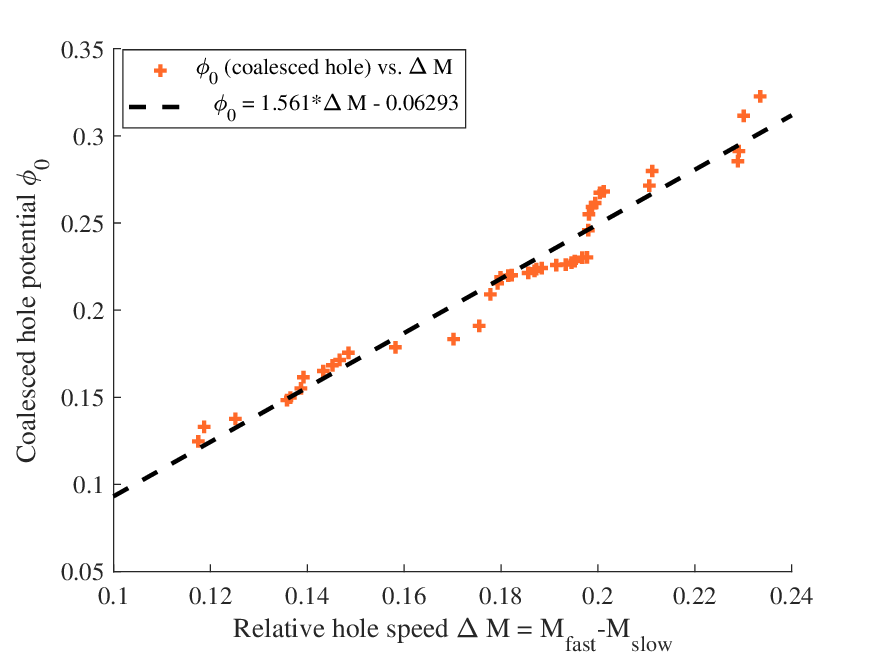}
    \includegraphics[width=0.48\linewidth]{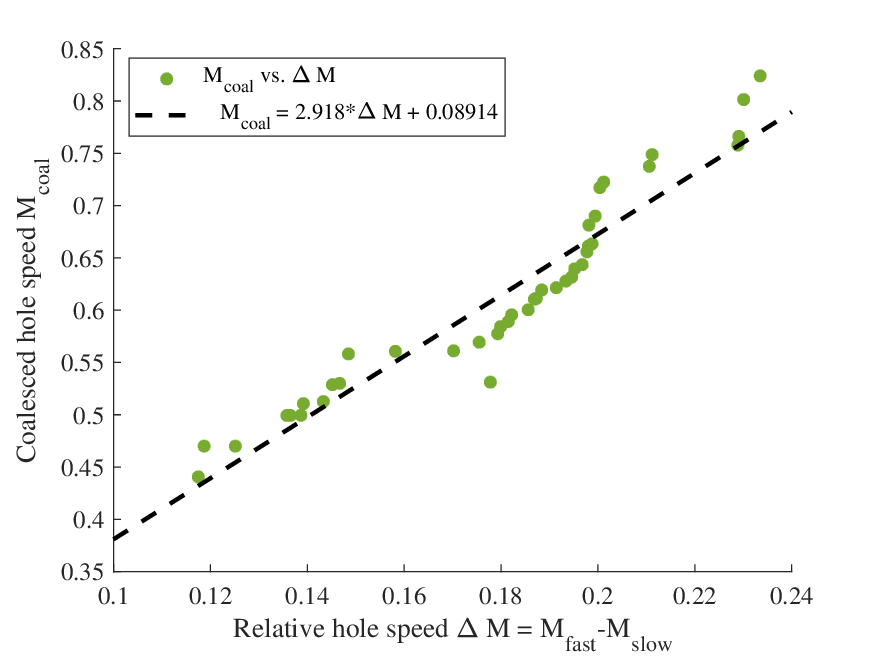}
    \caption{Variation of coalesced hole characteristics with relative speed of the colliding holes. Coalesced hole potential amplitude (left) and speed (right). Black dashed curves in each case representing approximated fittings.}
    \label{fig:coal_hole_characteristics}
\end{figure}

It was shown initially \cite{Lynov1980} that the coalescence of phase-space holes depends on the relative speeds of the holes. Specifically, it was shown that two holes coalesce only when their relative speeds $\Delta M$ is sufficiently small. In order to further explore the relation between the coalescence phenomena and the relative speed of the two colliding holes, we study the variations in the coalesced hole characteristics with changing relative hole speeds. These include the coalesced hole potential amplitude, spatial charge density amplitude and speed. These are shown in figure \ref{fig:coal_hole_characteristics}.

From figure \ref{fig:coal_hole_characteristics}(left and right) it is clear that the coalesced hole potential and speed bear an almost linear relation with the velocity of approach of the colliding holes, with a constant intercept as is visible.
\begin{equation}
    (\phi_0 - p_0) \propto \Delta M \Rightarrow \frac{\delta \phi_0} { \delta (\Delta M)} \approx \mathcal{C}.
    \end{equation}
    Here, $\delta$ represents a change in the quantities,  $p_0$ refers to the potential intercept, which from our linear regression we find to be approximately equal to $-0.06293 K_BT_ee^{-1}$. The constant $\mathcal{C}$ is found to be approximately equal to $1.5$, after normalisation. Similarly, 
    \begin{equation}
        ( M_{coal.} - m_0) \propto \Delta M \Rightarrow \frac{\delta M_{coal.}}{ \delta (\Delta M)} \approx \mathcal{M}.
    \end{equation}
Here, $m_0$ and $\mathcal{M}$ refer to the intercept and the slope, which we find from our linear fitting to be approximately equal to $0.09$ and $2.9$ respectively. 

\begin{figure}
    \centering
        \includegraphics[width=0.48\linewidth]{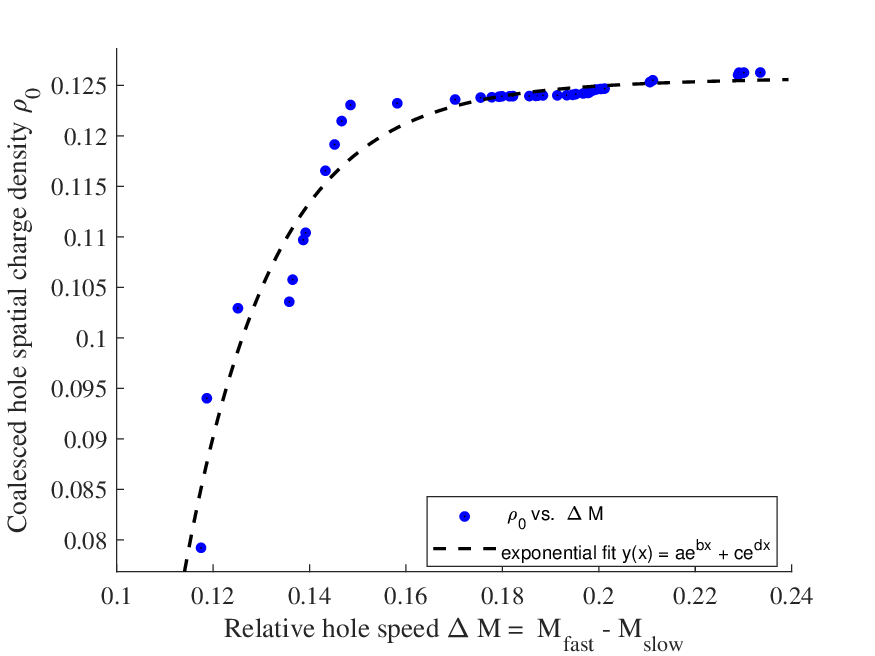}
    \includegraphics[width=0.48\linewidth]{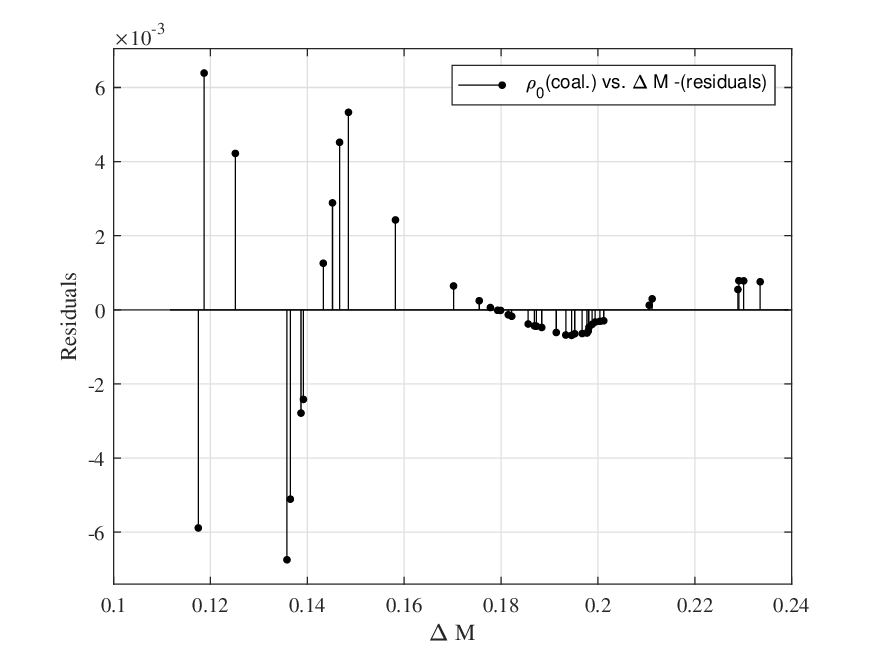}
    \caption{(Left) Variation of hole spatial charge density amplitude vs. relative hole speed of collision. (Right) Curve fitting residuals.}
    \label{fig:coal_hole_characteristics_rho}
\end{figure}
The spatial charge density amplitude of the coalesced hole seems to saturate with increasing values of $\Delta M$, as shown in figure \ref{fig:coal_hole_characteristics_rho}(left). In other words, as the relative speed of collisions increase, coalescence due to the hole collisions does not cause a growth in the spatial charge density. This is also in agreement to the observation presented earlier in figure \ref{fig:electron_hole_qm_potential_vs_amp}(top, right). It can be seen that as the pulse amplitude increases, the hole spatial charge density amplitudes do not continue to grow but instead seem to saturate. In order to approximate this relation, we have used exponential fitting of the form 
\begin{equation}
    \rho_0(\Delta M) = a\exp(b\Delta M) + c\exp(d\Delta M),
\end{equation}
where table \ref{tab:data_fitting_table} defines the fitting coefficients. The choice of the exponential fittings is in keeping with the analytical solutions of the hole charge density \cite{Schamel1979, Turikov1984}, which is in terms of sec-hyperbolic and tan-hyperbolic functions. We also find this fitting 
 presents the lowest residual values (shown in fig. \ref{fig:coal_hole_characteristics_rho}(right)), when compared to the polynomial fittings which do not present inconsistent relations outside the shown range. These results suggest that the hole relative speeds before collision bear a strong influence on the dynamics of the coalesced hole, apart from dictating the coalescence criteria.
\begin{table}
    \centering
    \begin{tabular}{|c|c|c|c|}
    \hline
        & Value & Lower & Upper \\
        \hline
       a & 0.1241 & 0.1069  & 0.1413 \\
       b & 0.0514 & -0.6135 &0.7163 \\
       c & -24.5122& -75.2819 &26.2575 \\
       d &-54.6744 &-73.3505  &-35.9983 \\
       \hline
    \end{tabular}
    \caption{Coefficients and 95\% Confidence Bounds for exponential curve fitting used for relating coalesced hole charge density amplitude $\rho_0$ with the relative hole speeds before their coalescence, $\Delta M$.}
    \label{tab:data_fitting_table}
\end{table}

\begin{figure}
    \centering
    \includegraphics[width=0.6\linewidth]{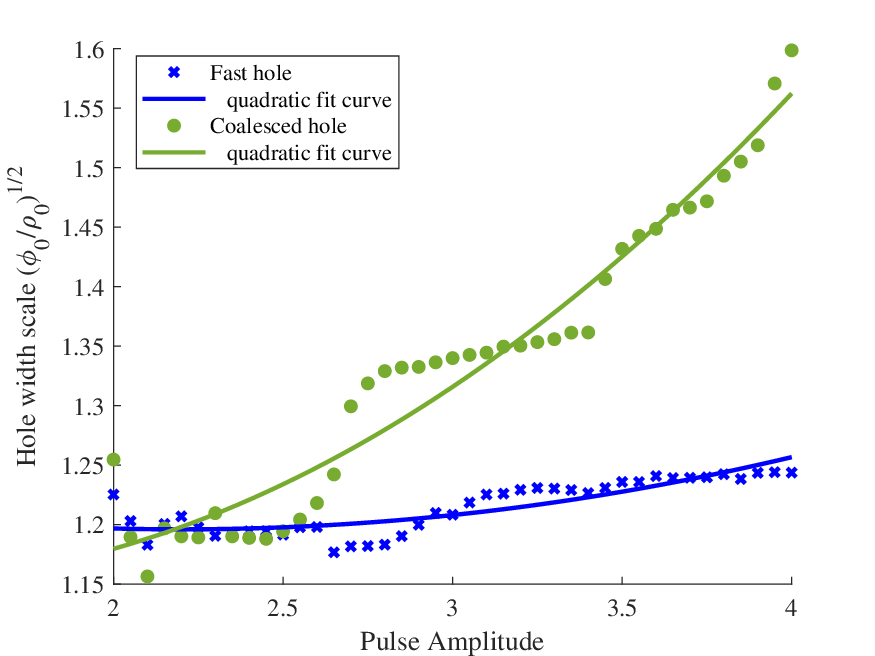}
    \includegraphics[width=0.35\linewidth]{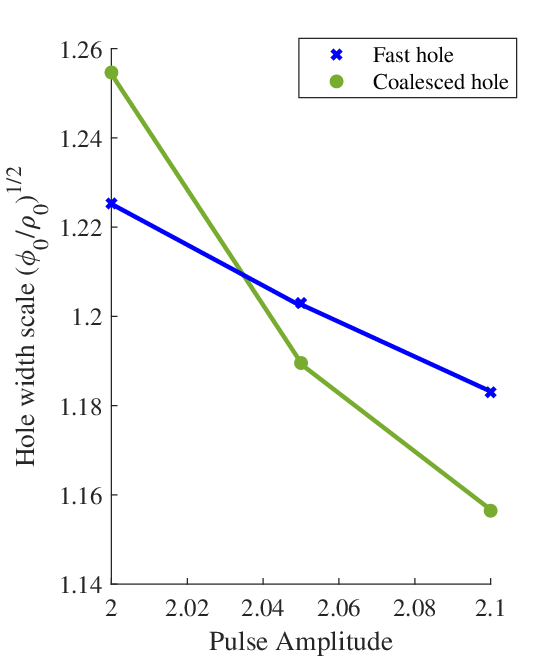}
    \caption{Hole spatial width scales for the coalescing and coalesced holes. For one of the colliding holes, the excitation amplitude is fixed at $3.0K_BT_ee^{-1}$, whereas the second hole excitation amplitude is $\mathcal{K}A$ where $\mathcal{K}$ is varied between $0.6$ and $1.3$. For each case, the hole spatial scale is measured according to equation (\ref{eqn:width}). Left panel depicts observations in the total range, whereas the right panel depicts the zoomed small amplitude case. }
    \label{fig:widthwidth}
\end{figure}
At the end of our investigation, we investigate the influence of coalescence on the hole spatial width. For this, we maintain the excitation amplitude for the first hole at $3K_BT_ee^{-1}$ and excite the second hole using $\mathcal{K}$ values between $[0.6, 1.3]$. The hole spatial width is estimated by obtaining the ratios of hole potential and charge density amplitude through the Poisson's equation:
\begin{equation}\label{eqn:width}
   \frac{\delta^2 \phi }{\delta x^2} = - \rho \Rightarrow \delta x \approx \sqrt{\phi_0/\rho_0}.
\end{equation}
We find that the coalesced hole characteristic length scale increases with the increase in the coalescing hole width scales. This is shown in fig. \ref{fig:widthwidth}. Upon correlating with individual hole amplitudes shown in figure \ref{fig:electron_hole_qm_potential_vs_amp}, it can be seen that for large amplitudes, the hole widths increase quadratically with the amplitudes. However for the small amplitudes, the relation is inversely proportional. This agrees to the hole width-amplitude dependence observed by Lynov et al. \cite{Lynov1985Phase-SpaceHoles} which links the theoretical predictions of Schamel \cite{Schamel1979, schamel1986electron} and that of Turikov \cite{Turikov1984} and Chen et al. \cite{Chen2005}.

\section{Conclusions}
In this work, we have intensively investigated the coalescence of phase-space holes in two plasma systems -- an infinite, two-stream plasma and a cylindrically wave-guided plasma. In the existing literature, these two set-ups have been ubiquitously employed for observations of the hole coalescence phenomena. In this study, we have explored the same by including parametric studies and the theories developed by the recent phase-space hydrodynamic approach.

It is shown that the phenomena of phase-space hole coalescence can be attributed, in the phase-space hydrodynamics analogy, to the interaction between the velocity fields of the phase-space vortices (holes). When in sufficiently close proximity to each other, these vortices in the plasma phase-space interact with each others' velocity fields, resulting in the fusion of the vortex cores and the resultant combined velocity field, which ultimately becomes a single hole. 

Parametric studies of the electron (and ion) coalescence in the cylindrically wave-guided plasma, which mimics the Q-machine experiment conducted earlier by Saeki et al. \cite{Saeki1979} and Lynov et al. \cite{Lynov1979}, reveal that the properties of the coalesced hole show many observable parametric relations with the individual properties of the colliding holes and their relative speeds. These results present a detailed outlook of the hole coalescence phenomena in the plasma set-ups employed in this study.

\section*{References}
\bibliography{references.bib}
\section*{Acknowledgement}
One of the authors (A.L.) acknowledges the TMA-PAI doctoral research fellowship provided by Sikkim Manipal University. 

\section*{Author Contributions}
V.K.S. has conceived the interaction hypothesis. Both A.L. and V.K.S. have conceived the experimental set-up. A.L. has developed the numerical code and conducted the simulation study. A.L. and V.K.S. analysed the results. A.L. has developed the theoretical framework and drafted this manuscript. Both  A.L. and V.K.S. have reviewed the manuscript. 

\section*{Additional information}
 \textbf{Competing interests:} The authors declare no competing interests. 

\noindent \textbf{Correspondence} and request for materials should be addressed to A.L.  

\bibliographystyle{ieeetr}
\end{document}